\documentclass[preprint]{elsarticle}
\usepackage{graphicx}
\usepackage{lscape}
\usepackage[margin=3cm, marginpar = 2cm]{geometry}
\usepackage{subfigure}
\usepackage{amsmath}
\usepackage[final]{changes}
\usepackage{marginnote}
\usepackage{color}
\numberwithin{equation}{section}

\definechangesauthor{ER}{orange}
\definechangesauthor{DS}{blue}
\begin{document}

\title{On the Stability of Metabolic Cycles}
\author[bnf]{Ed Reznik\fnref{fn1,fn2}}
\ead{ereznik@bu.edu}

\author[bnf]{Daniel Segr\`{e}\corref{cor1}\fnref{fn1,fn2,fn3}}
\ead{dsegre@bu.edu}

\cortext[cor1]{Corresponding Author}
\fntext[fn1]{Boston University Department of Biomedical Engineering}
\fntext[fn2]{Boston University Center for Biodynamics}
\fntext[fn3]{Boston University Department of Biology}
\address[bnf]{Bioinformatics Program, 24 Cummington St, Boston MA}

\begin{abstract}
We investigate the stability properties of two different classes of metabolic cycles using a combination of analytical and computational methods. Using principles from structural kinetic modeling (SKM), we show that the stability of metabolic networks with certain structural regularities can be studied using \added{a combination of analytical and computational techniques}. We then apply these techniques to a class of single input, single output metabolic cycles, and find \added{that the cycles are stable under all conditions tested}. Next, we extend our analysis to a small autocatalytic cycle, and determine parameter regimes within which the cycle is very likely to be stable. We demonstrate that analytical methods can be used to understand the relationship between kinetic parameters and stability, and that results from these analytical methods can be confirmed with computational experiments. In addition, our results suggest that elevated metabolite concentrations and certain crucial saturation parameters can strongly affect the stability of the entire metabolic cycle. We discuss our results in light of the possibility that evolutionary forces may select for metabolic network topologies with a high intrinsic probability of being stable. Furthermore, our conclusions support the hypothesis that certain types of metabolic cycles may have played a role in the development of primitive metabolism despite the absence of regulatory mechanisms.
\end{abstract}

\maketitle

\section{Introduction}
Cycles are at the heart of the metabolic networks of organisms spanning the entire tree of life \cite{ganti,morowitz,hypercycle}. For example, the tricarboxylic acid (TCA) cycle sits at the core of energy production for many species and additionally plays the role of regenerating essential cellular nutrients and components. A great deal of research has been devoted to understanding the stability properties of such cycles, which have often been represented as chemical reaction networks (CRNs). Many of these studies have drawn conclusions about the steady state of a CRN without regard to details about the rate constants or chemical concentrations of the particular CRN (for instance, see \cite{injective2,injective,CRNT}). Instead, these approaches rely on the topology of the network and commonly assume mass-action kinetics laws to limit (and sometimes determine) the possible behaviors the system may exhibit at steady state. Can one hope to extend these results to general, potentially nonlinear, rate laws?

In this work, we use a generalized modeling framework known as structural kinetic modeling (SKM) to extend previous results on two classes of metabolic cycles. The first class has a special but relatively flexible structure: it contains a single input, a single output, and unlimited length. This relatively organized structure makes it amenable to analytical methods of study. For cycles belonging to this first class, we prove \added{that the cycle can only lose dynamical stability in an oscillatory manner, regardless of absolute metabolite concentrations, flux values, kinetic rate constants, and form of monotonic kinetic rate law. We further investigate the cycle computationally, and find that it remains completely stable under all conditions tested, leading us to make a conjecture about its stability under all conditions.} 

We then apply the same SKM approach to a second class of cycles, autocatalytic cycles, and demonstrate that these are not stable under all combinations of kinetic parameters and metabolic conditions. Instead, we use analytical calculations to derive relationships between key parameters which can enhance or attenuate the stability properties of the cycle. These relationships are then computationally tested and verified. 

Nearly all prior published work using SKM \cite{SKM2,SKM3} employs computational experiments and statistical methods in order to draw conclusions about the system of interest. In this work, we show that metabolic systems with relatively well-organized structures are not only analytically tractable but furthermore that their analysis can actually lead to interesting and novel conclusions. Finally, we discuss the implications of our results on the evolution of primitive metabolic networks. The SKM technique used throughout the paper highlights essential parameters which directly impact the stability of the cycle under study. Particular combinations of these parameters may have conferred an evolutionary advantage to their respective metabolic networks by making them robust to small environmental perturbations.

\section{Background}
\subsection{Structural Kinetic Modeling (SKM)}
The methods developed in this paper are based on the SKM framework\cite{SKM}. SKM is a specific application of generalized modeling \cite{GM} in which normalized parameters replace conventional parameters (such as $V_{max}$ or $K_{M}$ in the modeling of metabolic networks). The normalized parameters have a direct connection to the original kinetic parameters, but are much easier to work with. As will be shown below, these parameters usually have well-defined and limited ranges (e.g. [0,1]), and sampling them across this range effectively samples all possible values of the original kinetic parameters.

The goal of SKM is to capture the local stability properties of a biochemical system. In this sense, it bridges the gap between genome-scale steady state modeling \cite{FBA} and explicit kinetic modeling of a metabolic process. To study stability, one usually determines the Jacobian $J$ of the system of interest and evaluates it at the system's steady states. Assuming knowledge of the form of kinetic rate laws for a CRN, it is quite possible to write down the corresponding $J$. However, in many cases $J$ will be unnecessarily complicated and quite difficult to work with. By performing a change of variables, SKM actually simplifies the mathematical form of each entry in $J$. The new entries are in almost all cases easier to work with. The specific method by which the change of variables occurs is briefly described below and more completely in \ref{sec:SKM}. Much of this material is based on \cite{SKM} and we refer readers to that reference and its supplementary materials for more information.

If we let \textbf{S} be the \textit{m}-dimensional vector of metabolite concentrations, $\mathbf{N}$ be the $\mathit{m} \times \mathit{r}$ stoichiometric matrix, and $\mathit{v}$ be the \textit{r}-dimensional vector of reaction rates, then we can describe the dynamics of the system with the equation

\begin{equation}
	\frac{d\mathbf{S}}{dt}=\mathbf{N\mathit{v}(S,k)}
\end{equation}

where $v(\mathbf{S,k})$ denotes that the reaction rates are dependent on both metabolite concentrations \textbf{S} and kinetic parameters (such as Michaelis-Menten constants) \textbf{k}. If we assume that a non-negative steady state $\textbf{S}^{0}$ exists, then we can redefine our system using the definitions

\begin{equation}	 \mathit{x_{i}}=\frac{\mathit{S_{i}(t)}}{S_{\mathit{i}}^{0}},\mathbf{\Lambda}_{ij}=N_{ij}\frac{\mathit{v_{j}}(\mathbf{S}^{0})}{S^{0}_{i}},\mu_{j}(\mathbf\mathit{x})=\frac{\mathit{v_{j}}(\mathbf{S})}{\mathit{v_{j}}(\mathbf{S}^{0})}
\end{equation}

where $i = 1...m$ and $j = 1...r$, $\mathbf{x}$ is a vector of metabolite concentrations normalized with respect to their steady state concentrations and $\mathbf{\mu}$ represents flux normalized with respect to steady state flux values. The matrix $\mathbf{\Lambda}$ represents the stoichiometric matrix normalized with respect to steady state fluxes and steady state metabolite concentrations.

The Jacobian, evaluated at $\mathbf{x}^{0} = 1$ (which, because of the way $\mathbf{x}$ is defined, is actually the equilibrium of the system), can be written as

\begin{equation}
\mathbf{J_{x}} = \mathbf{\Lambda \theta ^{\mu}_{x}}
\end{equation}

Note that the equations above were derived without regard to the actual form of the kinetic equations that determined the ODE system. The matrix $\mathbf{\theta^{\mu}_{x}}$ contains elements which represent the degree of saturation of normalized flux $\mathbf{\mu}_{j}$ with respect to normalized substrate concentration ${\mathbf{x}_{i}}$. In terms of derivatives, each element of $\mathit{\theta}$ represents the degree of change in a flux as a particular metabolite is incrementally increased. This is analogous to the concept of elasticity in metabolic control analysis \cite{MCA}.

What does the $\mathbf{\theta}$ matrix look like? Its columns correspond to each metabolite, and its rows to each flux. A non-zero element $\theta^{i}_{j}$ in the matrix represents the effect a small change in a metabolite $j$ has on flux $i$. In the case of Michaelis-Menten kinetics, this element in the matrix may take values ranging from [0,1]. In the case of standard competitive inhibition (e.g. allosteric inhibition by a product), the element takes values in [-1,0].

To usefully illustrate the meaning of a single $\theta$ element, consider an equation following Michaelis-Menten kinetics shown in (\ref{eq:MMapp}). First, we write $\mu(x)$, which we recall is the flux normalized by the flux at the steady state. Here, $S_0$ is the concentration of the substrate $S$ at steady state. Then, we manipulate the equation by substituting $xS_0$ for $S$, where $x$ is the normalized steady state concentration of substrate $S$. The result is shown in (\ref{eq:MMmu}).

\begin{equation}
V = \frac{k_{2}E_{0}S}{K_{M}+S}
\label{eq:MMapp}
\end{equation}

\begin{equation}
\mu(x) = \frac{\frac{k_{2}E_{0}S}{K_{M}+S}}{\frac{k_{2}E_{0}S_0}{K_{M}+S_0}} = x\frac{K_M + S_0}{K_M + xS_0}
\label{eq:MMmu}
\end{equation}

Finally, we take a derivative with respect to $x$ and evaluate it at $x = 1$ to obtain $\theta$ in (\ref{eq:MMderiv}). Notice that $\theta$ can only take values between 0 and 1 for any positive value of $S_0$.
\begin{equation}
\theta = \frac{1}{1+\frac{S_0}{K_M}}
\label{eq:MMderiv}
\end{equation}

The power of SKM comes from the parametrization illustrated above. Each element in $\mathbf{\theta}$ has a precise correspondence to some combination of kinetic parameters in the original model. However, it is far more tractable to study a system using $\theta$ parameters rather than the original kinetic parameters. To see this more clearly, recall that in most cases, biochemical kinetic constants are poorly estimated. In order to build precise ODE models for biochemical systems, it is usually necessary to actually choose values for these constants. While the chosen values may be estimated from experimental measurements, hidden dependencies in the model may actually result in non-obvious correlations between parameters that can strongly affect the output of the model.

On the other hand, if one is only interested in the \textit{stability} of a characterized steady-state, it may not be necessary to actually have knowledge of kinetic parameters. Experimental measurements can provide data on absolute metabolite concentrations and flux values, and the stoichiometry is often known \textit{a priori}. Then, one can parameterize the system as shown above and sample many possible combinations of $\theta$ parameters. For each unique set of $\theta$ parameters, the stability of the Jacobian is determined and recorded. Analysis of the results can lead to several important conclusions such as which $\theta$ parameters have the strongest correlation with stability of the entire metabolic system. Thus, the benefit of employing SKM over other techniques is that it provides the means to analyze and make sense of a large number of possible cases of a metabolic network, rather than a single instance.

At first glance, it is unclear whether this will make things easier or harder; values for fluxes and concentrations are almost never known. However, our work to follow shows that the absence of this knowledge does not make the problem of stability intractable. Instead, we operate with the following idea: we assume that the steady state flux vector $v$ and the metabolite concentration vector $X$ are variables, and we do not specify them. Instead, we simply determine the stability of the system in terms of these and other ($\mathbf{\theta}$) variables. Then, we can find trends in stability as the flux and metabolite concentrations change: it may be that as a concentration goes up, the stability of the system tends to fall. Isolating these trends is the motivation of our analysis.

\subsection{Previous results on stability of metabolic cycles}

Our work is preceded by a great deal of work aimed at understanding the stability of metabolic networks\cite{hypercycle,doyle_autocat,king, king2, tyson}. One flavor of such work is referred to as chemical reaction network theory (CRNT). CRNT is an elegant and powerful framework for understanding how the structure of chemical networks dictates their equilibrium and stability properties. By applying CRNT to the cycles studied in this paper, one can \added{(in some cases)} prove the stability of their steady states. This result can be easily obtained, for example, by feeding the stoichiometric matrices of the corresponding networks into the ERNEST MATLAB Toolbox \cite{ERNEST}. Notably, these results cannot be easily extended beyond the case of mass-action kinetics (where the rate of reaction is proportional to the concentration of its reactants). The SKM method presented in this article extends the stability results for these cycles to general \added{monotonic} kinetics.

{Dynamical systems approaches have also been used in understanding the stability properties of autocatalytic cycles (the second type of cycle presented in the paper). Previous research by King \cite{king} (focusing on dynamics) and by Eigen \cite{hypercycle} (focusing on steady-state stability properties) have studied autocatalytic cycles assuming first-order (non-saturating) mass-action kinetic rate laws. Extending such work, our analysis takes into account all forms of saturating rate laws up to order one, and derives simple expressions for the stability of metabolic cycles given adequate information about rate constants and steady-state concentrations. To put this in context, the prior work on autocatalytic cycles represents a single point (at which all $\theta$ parameters equal 1) in the large space of $\theta$ parameters within which the actual metabolic network resides. Our approach searches over this entire space, and identifies trends in stability across it.

\section{A Simple Example}
To illustrate the main concepts behind analytically characterizing the stability of a metabolic cycle using SKM, we will start with a simple example. Consider the cycle shown in Figure \ref{fig:simple}. To analyze the conditions under which it is stable, we will write the $\mathbf{\Lambda}$ and $\mathbf{\theta}$ matrices, multiply them together to obtain the Jacobian $J$, and then find the eigenvalues of $J$ by writing out its characteristic equation. 

\begin{figure}[h!]
	\begin{center}
		\includegraphics{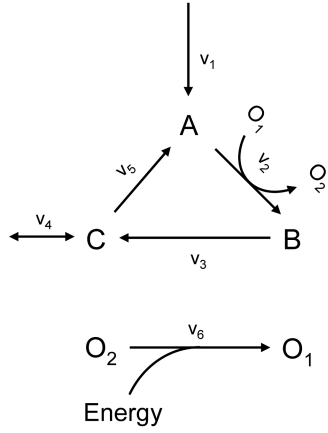}
	\end{center}
	\caption{A simple metabolic cycle}
	\label{fig:simple}
\end{figure}

The cycle contains 3 metabolites and 6 reactions. We impose that the reactions can only proceed in the forward direction. Furthermore, the 6 reactions are constrained by mass balance, and we note that two linearly independent rate vectors will characterize the fluxes of all 6 reactions in the network. Assume that the steady states metabolite concentrations $(A^0,B^0,C^0,O_1^0,O_2^0)$ of the system are $(1,1,1,1,1)$. Furthermore, assume that a flux of magnitude $\alpha F$ enters the cycle through reaction $v_{1}$, and $(1-\alpha)F$ flux returns through reaction $v_{5}$ (where $0 < \alpha < 1$).

Finally, we impose that there is no complicated activation or inhibition present in the system. This means that $v_2$ uses only metabolite $A$ and cofactor $O_1$ as its substrates, $v_3$ uses only metabolite $B$, $v_4$ and $v_5$ use only metabolite C, $v_6$ uses only $O_2$, and $v_1$ is constant and uses none of the metabolites as substrates. Notice that this constrains all $\theta$ parameters to be greater than or equal to zero. \added{Constraining $\theta$ to be positive ensures that any kinetic rate laws we consider in our analysis are monotonic, i.e. that an increase in the concentration of the substrate of a reaction will never decrease the rate of a reaction.} This will be essential in the analysis to follow. Now, we can write our system in terms of SKM variables:

\begin{equation}
\Lambda =
\left[
\begin{array}{cccccc}
\alpha F & -F & 0 & 0 & (1-\alpha) F & 0\\
0 & F & -F & 0 & 0 & 0\\
0 & 0 & F & -\alpha F & -(1-\alpha) F & 0\\
0 & 0 & -F & 0 & 0 & F\\
0 & 0 & F & 0 & 0 & -F
\end{array}
\label{eq:lamb_simple}
\right]
\end{equation}

\begin{equation}
\theta =
\left[
\begin{array}{ccccc}
0 & 0 & 0 & 0 & 0\\
\theta_{1} & 0 & 0 & 0 & 0 \\
0 & \theta_{2} & 0 & \theta_5 & 0\\
0 & 0 & \theta_{3} & 0 & 0\\
0 & 0 & \theta_{4} & 0 & 0\\
0 & 0 & 0 & 0 & \theta_6
\end{array}
\right]
\end{equation}

Because the cofactors $O_1$ and $O_2$ are conserved, we can remove them from the two matrices above (for details and justification, see the Supplementary Material of \cite{SKM}). Doing this, we omit the last row of $\mathbf{\Lambda}$, and $\mathbf{\theta}$ becomes
\begin{equation}
\theta =
\left[
\begin{array}{ccccc}
0 & 0 & 0 & 0\\
\theta_{1} & 0 & 0 & 0\\
0 & \theta_{2} & 0 & \theta_5\\
0 & 0 & \theta_{3} & 0\\
0 & 0 & \theta_{4} & 0\\
0 & 0 & 0 & -\theta_6
\end{array}
\right]
\label{eq:theta_simple}
\end{equation}

Then, to obtain the Jacobian we compute the product $\mathbf{\Lambda \theta}$. In order to determine the stability of the system, we find the eigenvalues of the Jacobian and determine whether any of them are positive. We find the eigenvalues by imposing $|J - \lambda I|=0$, i.e. by setting to zero the following determinant:

\begin{equation}
\left|
\begin{array}{cccc}
-\lambda - F\theta_{1} & 0 & (1-\alpha )F\theta_{4} & 0\\
F\theta_{1} & -\lambda -F\theta_{2} & 0 & -\theta_5\\
0 & F\theta_{2} & -\lambda -\alpha F\theta_{3} -(1-\alpha)F\theta_{4} & \theta_5 \\
0 & -\theta_2 & 0 & -\theta_5 - \theta_6
\end{array}
\right|
\end{equation}

This corresponds to the equation

\begin{eqnarray}
(-\alpha\theta_3-(1-\alpha)\theta_4 - \lambda)(\theta_1-\lambda)\left( (-\theta_5-\theta_6 - \lambda)(-\theta_2-\lambda) - \theta_2\theta_5 \right) + \\
(1-\alpha)\theta_1\theta_2\theta_4(-\lambda-\theta_6) = 0
\end{eqnarray}

The stability properties will be determined by the signs of the real parts of the eigenvalues obtained from solving this equation. \added{For all eigenvalues with only real components}, it is possible to determine the signs without explicitly solving the equation, by using Descartes' Rule of Signs. If we arrange the characteristic polynomial in order of decreasing variable exponents, then the number of positive \added{real} roots is equal to the number of signs changes in the coefficients of the ordered polynomial, or less than the number of size changes by a multiple of 2. In our case, there are no sign changes (all coefficients are greater than zero, notice that any negative terms cancel), so there are no positive roots. Note that it is not possible to have a zero eigenvalue either, because there is a nonzero coefficient for the $\lambda^{0}$ term. \added{Then, we must consider two alternative cases. In one, the Jacobian has only real eigenvalues, in which case they are all negative and the system is stable. In the other, the Jacobian may contain pairs of complex conjugate eigenvalues whose real part will determine the stability of the system. To investigate systems which may display such complex eigenvalues, we sampled many instances of the Jacobian computationally and computed the corresponding eigenvalues. Any eigenvalues with real part greater than zero would indicate instability. For this particular cycle, we randomly sampled each $\theta$ in the range (0,1] for a value of $\alpha$ in the open set (0,1). Then, the eigenvalues of the Jacobian were calculated. This process was repeated $4 \times 10^6$ times and in no cases were any unstable Jacobians identified.}

\section{A More General Treatment}
In this section we generalize our previous results to a metabolic cycle of arbitrary length, relying on the following assumptions:
\begin{enumerate}
	\item There exists at least a single steady state for the system. It is trivial to show that such a steady state exists because rate constants (such as $V_{max}$) are not related to any $\mathit{\theta}$ parameters which determine the stability, and can be chosen independently and arbitrarily in order to satisfy the steady-state condition. Furthermore, results from \cite{injective2} and computational analysis using the ERNEST MATLAB package \cite{ERNEST} demonstrate that the cycle under study is strongly sign determined (SSD) and injective. It therefore admits only a single unique equilibrium for a given set of kinetic parameters (e.g. rate constants, saturation constants, etc.). It is worthwhile to note that if multiple equilibria did exist, the results presented in this section would still be valid.
	\item The input and output reactions from the cycle are separated by one metabolite. In other words, if the output reaction of the cycle leaves through metabolite $x_{i}$ in the cycle, then the input reaction to the cycle enters through metabolite $x_{i+1}$. In analogy to the simple example above, the output reaction leaves through metabolite $C$, and the input reaction enters in the next metabolite in the chain, $A$. This assumption is made to simplify the analysis, but computational experiments indicate it does not appear to be necessary. However, this has not been explicitly proven.
	\item The $j^{th}$ reaction ($j = 2,3,...,N+1$) uses only the $(j-1)^{th}$ metabolite as a substrate. There is no feedback inhibition or activation of any kind. This means that all $\theta$ parameters are constrained to be greater than zero. A negative $\theta$ would correspond to a metabolite having an inhibitory effect on the rate of reaction.
	\item Cofactors such as ATP and NADH are often used as the energetic driving forces in metabolic reactions in order to make these reactions thermodynamically feasible. We assume that a single cofactor pair ($O_1$ and $O_2$) is involved in the metabolic cycle, \added{and that it is not involved in the reaction creating the first or last metabolite in the cycle}. The reformation of cofactors is assumed to be driven by some energy input into the system.
	
\end{enumerate}

Our goal is to \added{study} the stability of the cycle shown in figure \ref{fig:bigcyc}, which generalizes the cycle of Figure \ref{fig:simple}.

\begin{figure}[h]
	\begin{center}
		\includegraphics{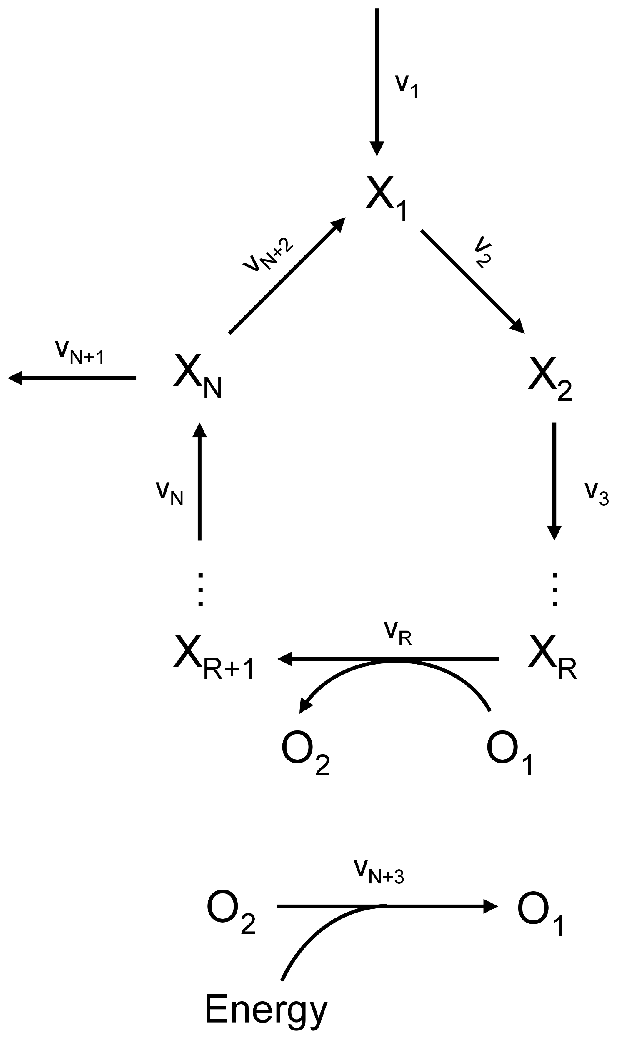}
	\end{center}
	\caption{A complete metabolic cycle with cofactors}
	\label{fig:bigcyc}
\end{figure}

The generalized form of $\mathbf{\Lambda}$ and $\mathbf{\theta}$, analogous to the ones from (\ref{eq:lamb_simple}) and (\ref{eq:theta_simple}), is shown in Appendix \ref{sec:Cofactors}. Note that we make no further assumptions about the steady states of the system: $X^0 = (X_1^0, X_2^0,\dots,X_N^0)$ and the flux through the system remains equal to $F$. With $N$ metabolites and $N+3$ reactions, $\Lambda$ is $N \times (N+3)$ and $\theta$ is $(N+3) \times N$. The complete analysis is shown in \ref{sec:Cofactors}, and the results are summarized below.

The characteristic equation for the eigenvalues of the system reduces to

\begin{eqnarray}
\left( \frac{-\theta_{N+2}}{O_1^0} - \frac{\theta_{N+3}}{O_2^0} - \lambda \right) \left( \prod_{i=1}^{N-1} { \left(-\lambda - \frac{F\theta_i}{X_i^0}\right)} \left(-\lambda - \frac{ \alpha \theta_N + (1- \alpha) \theta_{N+1}}{X_N^0} \right)\right)+ \nonumber \\
\left( \frac{-\theta_{N+2}}{O_1^0} - \frac{\theta_{N+3}}{O_2^0} - \lambda \right) \left((-1)^{N-1}\frac{(1-\alpha) \theta_{N+1}}{X_1^0}\prod_{i=1}^{N-1}\frac{\theta_i}{X_{i+1}^0} \right) + \nonumber \\
(-1)^{2N+2R+1}\frac{\theta_R \theta_{N+2}}{O_1^0X_R^0}\left( \frac{-\alpha \theta _N - (1-\alpha )\theta_{N+1}}{X_N^0} - \lambda \right)\prod^{N-1}_{i=1,i\neq R} {\left(\frac{-\theta_i}{X_{i}^0}-\lambda \right)} + \nonumber \\
(-1)^{2N+2R+3} \frac{\theta_{R} \theta_{N+2}}{O_1^0 X_{R+1}^0}(-1)^{N}\frac{(1-\alpha)\theta_{N+1}}{X_1^0}\prod^{N-1}_{i=1,i\neq R}{\frac{\theta_i}{X_{i+1}^0}} = 0
\end{eqnarray}

The terms on the bottom two lines will cancel with terms from the top two lines, resulting in a polynomial equation of order $N+1$ containing all nonzero coefficients with exactly the same sign. Furthermore, a constant $\lambda^0$ term will remain, indicating that there is no zero eigenvalue. Therefore, all eigenvalues \added{with only real part are negative, and the system can only lose stability in an oscillatory manner}.

\added{We proceeded with computational experiments in which SKM parameters were randomly sampled to search for cases of systems whose eigenvalues lie in the positive real part of the complex plane. In these experiments, we generated random samples of $\mathit{\theta}$ parameters (in the range (0,10] to account for high Hill coefficients), $\alpha$ in (0,1), metabolite concentrations in (0,100), and calculated the eigenvalues of the Jacobian. Approximately $4 \times 10^6$ different instances were simulated for each particular cycle length from 3 to 8, and for each possible cofactor location. In none of the sampled instances was the real part of any eigenvalue of the Jacobian greater than zero, indicating that all samples were stable.} 

In a separate computational experiment, we tested the dependency of our stability results on the structure of the cycle. An example is shown in the lower part Figure \ref{fig:hist} in which the first metabolite activates the reaction creating the third metabolite \added{(similar to a feedforward loop)}. This introduces an instability into the system which does not appear in our prototypical system. 

\added{Thus, both computational experiments and analytical results support the conjecture that under general kinetic rate laws, cycles of this form are stable under all conditions. We believe that a proof of this conjecture may follow from methods similar to those used to establish stability results for linear chains with positive feedback, as presented in \cite{tyson}.}

\begin{figure}[h!]
	\centering
	\subfigure[]{
		\includegraphics{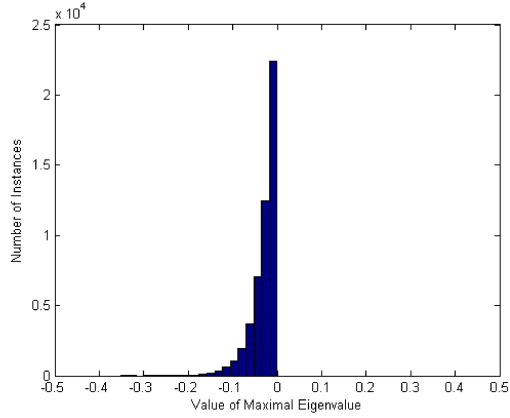}
	\label{fig:hist_stable}
	}
	\subfigure[]{
		\includegraphics{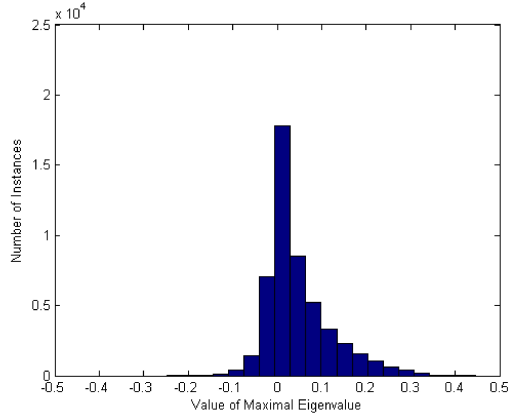}
	\label{fig:hist_unstable}
	}
\caption{\subref{fig:hist_stable} Histogram of values of the maximal eigenvalue for a case of the system shown in Figure \ref{fig:bigcyc}. The specific system contains 6 metabolites, and the reaction $v_2$ is coupled to a cofactor. All eigenvalues are stable, indicating the system is stable. \subref{fig:hist_unstable} Histogram of values of the maximal eigenvalue for a system in which one of the assumptions is broken. In this case, metabolite $X_1$ activates reaction $v_3$. Notice that not all maximal eigenvalues are negative, indicating that the system may be unstable for certain combinations of parameters. This suggests that global stability strongly relies on the topology of the network, and that additional activating or inhibitory reactions may prevent a steady state from being stable.}
\label{fig:hist}		
\end{figure}

\section{Autocatalytic Cycles}

The next question we want to address is whether the methods and results we derived for the cycles of metabolic reactions seen in the previous sections can be extended to cycles of fundamentally different topology, namely autocatalytic cycles. Autocatalytic cycles consist of a set of metabolites which reproduce themselves with each turn of the cycle. Their simplified stoichiometry often takes the form $A + X \rightarrow 2A$, where a substrate $X$ transforms into a product $A$, through catalysis of $A$ itself \cite{ganti,blackmond}. The formose reaction and the reductive citric acid cycle are two cases of autocatalytic cycles; both have been hypothesized to play roles in early metabolism, and significant work has been devoted to understanding the conditions under which they and other autocatalytic cycles may have naturally appeared and proliferated \cite{morowitz,protocell}.

Shown in Figure \ref{fig:autocat}, our simple example contains 3 metabolites and 5 reactions. It is distinct from the cycles shown in prior sections because both the first and last metabolite ($A$ and $C$) are required for reaction $v_2$ to occur. Furthermore, we assume that reactions exiting from the cycle transport metabolites to a separate compartment, where they are not accessible to the reactions within the cycle. Notably, we will show that this system does not contain a universally stable steady state. Instead, the steady state remains stable under certain parameter regimes, to be defined in the analysis below. Although this example is relatively simple, the results can be extended to more complicated examples using a method similar to the one described in Appendix C.

\begin{figure}[h!]
	\centering
	\includegraphics{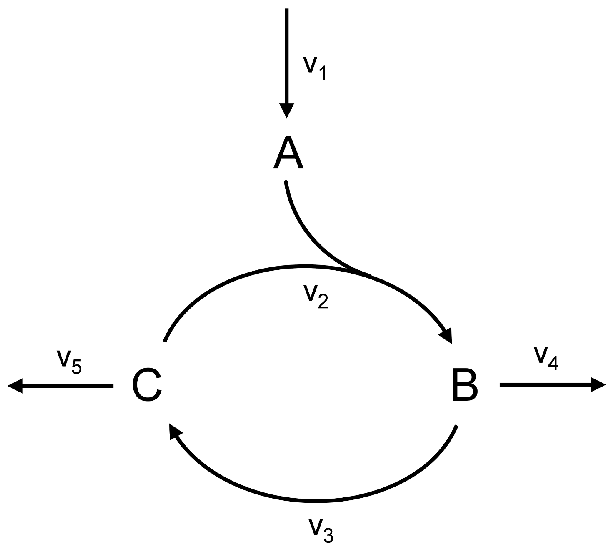}
	\caption{A simple autocatalytic cycle}
	\label{fig:autocat}
\end{figure}

The same assumptions are made as before, and all reactions are assumed to proceed in only the forward direction. There are 2 independent flux vectors which span the null space of the stoichiometric matrix for this system. Therefore, we assign the flux vector $V = (v_1,v_2,v_3,v_4,v_5) = (\alpha F,F, F(x - \gamma),\gamma F, F(x-\gamma-1))$, where $\gamma F$ is the outflow from the cycle through the reaction $v_4$ and $F(x-\gamma-1)$ is the outflow from the cycle through the reaction $v_5$.  Furthermore, it is noted that for flux balance considerations, $x>\gamma$. Since all reactions are assumed to go in the forward direction, $x>\gamma+1$.

Following the same analysis as before, we can write the matrices $\mathbf{\Lambda}$ and $\mathbf{\theta}$:
\begin{equation}
\Lambda = F
\left[
\begin{array}{ccccc}
\frac{1}{A_0} & \frac{-1}{A_0} & 0 & 0 & 0 \\
0 & \frac{x}{B_0} & \frac{-(x-\gamma)}{B_0} & \frac{-\gamma}{B_0} & 0\\
0 & \frac{-1}{C_0} & \frac{x-\gamma}{C_0} & 0 & \frac{-(x-\gamma -1)}{C_0}
\end{array}
\right]
\end{equation}

\begin{equation}
\theta =
\left[
\begin{array}{ccc}
0 & 0 & 0 \\
\theta_{1} & 0 & \theta_4\\
0 & \theta_{2} & 0\\
0 & \theta_{3} & 0\\
0 & 0 & \theta_{5}
\end{array}
\right]
\end{equation}

Then, the Jacobian $J$ is the product of $\mathbf{\Lambda}$ and $\mathbf{\theta}$:
\begin{equation}
\theta = F
\left[
\begin{array}{ccc}
\vspace{5pt}
\frac{-\theta_1}{A_0} & 0 & \frac{-\theta_4}{A_0} \\
\vspace{5pt}
\frac{x\theta_{1}}{B_0} & \frac{-(x-\gamma)\theta_{2} - \gamma \theta_3}{B_0} & \frac{x\theta_4}{B_0}\\
\vspace{5pt}
\frac{-\theta_1}{C_0} & \frac{x-\gamma \theta_{2}}{B_0} & \frac{-\theta_4 - (x-\gamma-1)\theta_5}{C_0}
\end{array}
\right]
\end{equation}

By imposing $|J-\lambda I|=0$, we get the characteristic equation
\begin{eqnarray}
\left( \frac{-(x-\gamma)\theta_{2} - \gamma \theta_3}{B_0} - \lambda \right) \Bigg( \left( \frac{-\theta_1}{A_0} - \lambda \right) \left(\frac{-\theta_4 - (x-\gamma-1)\theta_5}{C_0} - \lambda \right) - \frac{\theta_1 \theta_4}{A_0C_0} \Bigg) \nonumber \\
+\left(\frac{x(x-\gamma) \theta_{2} \theta_4}{B_0C_0} \right) = 0
\label{eq:autocat_char}
\end{eqnarray}

Expanding this equation, the coefficients of the third-order polynomial in $\lambda$ are:
\begin{align*}
 & \lambda^3 : 1 \\
 & \lambda^2 : \frac{\theta_1}{A_0} + \frac{\theta_4 + (x-\gamma-1)\theta_5}{C_0} + \frac{(x-\gamma)\theta_{2} + \gamma \theta_3}{B_0} \\
 & \lambda^1 : \frac{(x-\gamma)\theta_2}{B_0} \left(\frac{\theta_1}{A_0} + \frac{\theta_4 (1-x)}{C_0} \right) + \frac{(x-\gamma-1)\theta_5}{C_0} \left(\frac{\theta_1}{A_0} + \frac{(x-\gamma)\theta_2}{B_0} + \frac{\gamma \theta_3}{C_0} \right) + \frac{\gamma \theta_3}{B_0} \left(\frac{\theta_1}{A_0} + \frac{\theta_4}{C_0} \right) \\
  & \lambda^0 :  \left(\frac{(x-\gamma)\theta_{2} + \gamma \theta_3}{B_0} \right) \left(\frac{(x-\gamma-1)\theta_1 \theta_5}{A_0C_0}\right)
\end{align*}

To analyze the signs of the roots of this polynomial, we use the Routh-Hurwitz criterion. According to this criterion, a third-order polynomial in $z$ with coefficients, $a,b,c$ of the form $z^3 + az^2 + bz + c= 0$ has roots \added{lying in the negative real half of the complex plane} if the conditions $a > 0, c > 0, ab - c > 0$ are satisfied. For our polynomial, it is obvious that both $a$ and $c$ are always greater than zero as long as $x > \gamma + 1$, so we are left to find conditions where the last inequality is satisfied. Note that if $x < \gamma + 1$, then $c<0$, there is a positive root to the characteristic equation, and the system is unstable.

Rather than deriving a precise condition for stability in terms of any single variable, we can use simple observations regarding the particular forms of the coefficients to infer trends in stability. In particular, notice that all negative terms in $-c$ are precisely canceled out by corresponding terms in $ab$. Then, the only term which may cause $ab-c$ to be less than zero is $\frac{(x-\gamma)\theta_{2} \theta_4 (1-x)}{B_0C_0}$. There are several ways to minimize the negative value of this term, such as: decreasing $x$ (while ensuring that $x > \gamma + 1$), increasing $\gamma$ (while ensuring that $x> \gamma + 1$), decreasing $\theta_2$ and/or $\theta_4$, increasing $\theta_1$ and/or $\theta_3$ and/or $\theta_5$, decreasing $A_0$, and increasing $B_0$ or $C_0$.

We can computationally test these predictions by following a procedure similar to \cite{SKM} and randomly sampling the parameters that define the autocatalytic cycle. Specifically, we can calculate the eigenvalues of the Jacobian matrix as a single parameter of interest (such as a $\theta$ parameter) is varied, while all other $\mathbf{\theta}$ parameters and metabolite concentrations are sampled randomly. By doing so, we sample over many possible instances of the metabolic network, and can identify correlations between individual parameters and stability. For each combination of dependent variable (e.g.: $x,\theta_1,\theta_2,A_0$) and $\gamma$ shown in the figures below, $5 \times 10^5$ instances of the Jacobian were sampled and characterized. The value of the maximal eigenvalue for each instance indicated whether the network is stable or not.

From the terms of the $\lambda^0$ coefficient derived above, one can see that as long as $x>\gamma+1$, it is ideal for $\gamma$ to be as large as possible and $x$ to be as small as possible. This minimizes the value of $\frac{(x-\gamma)\theta_{2} \theta_4 (1-x)}{B_0C_0}$. In Figure~\ref{fig:vx}, the stability of the system is plotted against values of the stoichiometric coefficient $x$ of metabolite $B$ in reaction $v_2$. Each line indicates a different value for the outflow $\gamma$. As $x$ increases, the stability of the system falls. Additionally, as the value of $\gamma$ increases, the stability of the system increases, while the general shape of the plot remains identical. Note that if $x<\gamma+1$, then the system is universally unstable, in agreement with the condition derived before. It is particularly interesting that the system's optimally stable operating point is located at the boundary of a region where it is entirely unstable due to dilution effects. 

\begin{figure}[h!]
	\centering
	\subfigure[]{
		\includegraphics{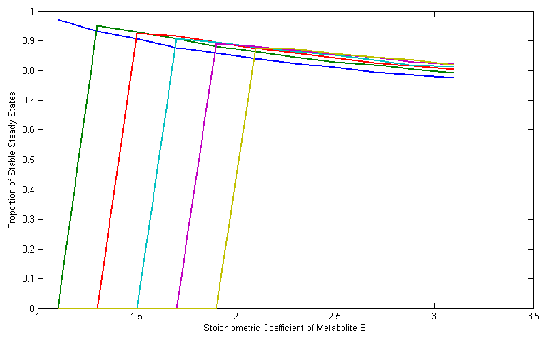}
	\label{fig:vx}
	}
	\subfigure[]{
		\includegraphics{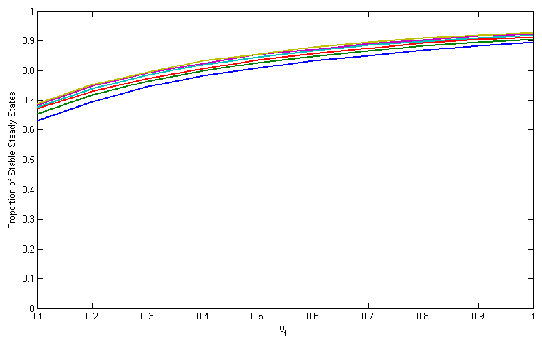}
	\label{fig:vt1}
	}
	\subfigure[]{
		\includegraphics{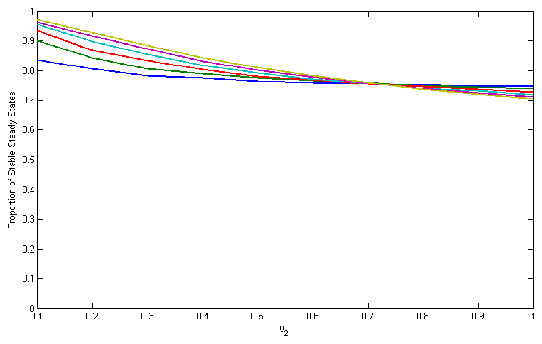}
	\label{fig:vt2}
	}
	\subfigure[]{
		\includegraphics{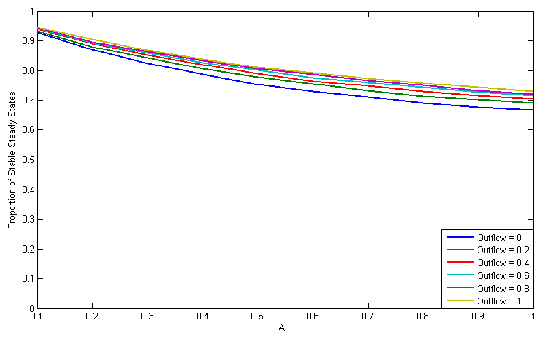}
	\label{fig:va}
	}
\caption{Each point in the figure is determined by sampling the Jacobian $5 \times 10^5$ times.  \subref{fig:vx} Proportion of stable steady states versus stoichiometric coefficient $x$, for different values of $\gamma$, \subref{fig:vt1} versus $\theta_1$, for different values of $\gamma$, \subref{fig:vt2}  versus $\theta_2$, for different values of $\gamma$, \subref{fig:va}  versus $A_0$, for different values of $\gamma$}
\end{figure}

Similarly, Figure~\ref{fig:vt1} confirms that as $\theta_1$ increases, the stability of the system rises. In Figure~\ref{fig:vt2}, it is shown that as $\theta_2$ increases, the stability of the system falls. Furthermore, the relationship between stability and $\gamma$ is retained in all these plots.

The combination of analytical and computational results presented so far in this section indicates that the tuning of a single parameter can greatly increase the likelihood of a metabolic system being stable. One may wonder which parameters, among those that can affect stability, are the ones that a cellular system would be most likely to adjust.  An intriguing possibility supported by the current findings is that a cell could approach a stable steady state by appropriately tuning metabolite concentrations. A sufficiently high steady state concentration of $B$ or $C$ would nearly guarantee that the simple autocatalytic cycle would be stable. In the future it may be interesting to verify experimentally whether certain metabolic pathways - and autocatalytic cycles in particular - may display steady state metabolite concentrations that fall into the range of high stability.

\section{Discussion}
We have used structural kinetic modeling (SKM) to assess the dynamical stability of certain classes of metabolic cycles. In prior work, it was demonstrated that SKM can be used as a powerful computational tool for identifying stabilizing sites in metabolic networks by sampling large ranges of parameters \cite{SKM2}. The main thrust of our work was to show that SKM can also be used analytically to characterize stability. Drawing from basic concepts in nonlinear dynamics and linear algebra, we demonstrated that it is possible to make a strong assertion about the stability of a class of non-autocatalytic metabolic cycles, and to identify trends in stability as a function of model parameters for a type of autocatalytic cycles.

For the non-autocatalytic cycles of the type illustrated in Figure \ref{fig:bigcyc}, we have proven that the topology of the networks ensures \added{that they can only lose stability in an oscillatory manner, regardless of metabolite concentrations, flux magnitude, or form of kinetic equations. The generality of the result with respect to kinetic laws is particularly noteworthy. Computational experiments support the conjecture that these cycles are completely stable and do not produce Jacobians with positive complex eigenvalues.} The stability of non-autocatalytic cycles similar to the one studied in this section could be easily proven using the Deficiency One Theorem of Chemical Reaction Network Theory (CRNT), \cite{CRNT}. However the CRNT result, in this case, would be restricted to cases of simple mass action kinetics. Hence, our work shows that an analytical approach completely distinct from CRNT converges to the same strong result, and extends it to a large category of kinetic rate laws. By approaching the question of stability with an analytical flavor of SKM, we demonstrate that the dynamics of a well-organized metabolic network, even in the presence of nonlinear kinetic rate laws, is not necessarily beyond ``pencil and paper'' methods, and that useful and significant conclusions can be drawn from such analyses. Additionally, recent work has shown that a broad category of regulatory and biochemical networks can be analyzed using the theory of monotone systems \cite{Sontag}. In the future it will be interesting to determine whether this theory applies to the cycles discussed here, and conversely how our analytical use of SKM may tie into the theory of monotone systems.

In applying the same SKM techniques to a simple autocatalytic cycle (Figure \ref{fig:autocat}), we found that a combination of analytically derived conditions and computational experiments could highlight the dependence of stability on metabolite concentrations, fluxes, and degrees of saturation. In this case, the stability of the cycle was not guaranteed, and the analysis suggested possible directions that evolution may take in order to steer the network towards an even more robust steady state. While it was possible to derive expressions equivalent to Eq.(\ref{eq:autocat_char}) without employing the SKM non-dimensionalization, it should be obvious that the algebra required to do so would be complicated and non-trivial assuming the rate laws were nonlinear. SKM provided an appealing (but nevertheless rigorous and useful) workaround, and still produced the correct characteristic equation for the Jacobian. To date, we are not aware of a similar effort to characterize the dynamic properties of autocatalytic cycles for general kinetic rate laws.

Our results suggest that a system does not necessarily require complex forms of regulation in order to maintain a stable steady state. In fact, in many cases it seems plausible that the elevated concentration of a single metabolite may be sufficient to endow stability. Why is it then that we observe many complex forms of allosteric (and genetic) enzyme regulation among biochemical networks? Certainly, such regulation may play a crucial role in stabilizing more complex enzymatic networks, but perhaps there is more to the story. One possibility may be that these forms of regulation actually control the transition of the system from one steady state to another. When conditions in the environment change, it may be favorable for a cell to alter its metabolism to react to these changes. Allosteric and transcriptional regulation may determine the manifolds along which the transitions between two steady states actually occur.

It is also interesting to consider the possibility that the inherent stability of certain types of pathways may influence the evolution of biochemical networks. One possible link between dynamic stability and evolution might be the adaptive fine tuning of kinetic parameters to guarantee stability of a given pathway topology. Conversely, we are suggesting that natural selection might gradually amplify the preference for network topologies with an intrinsically high probability of being stable. Such topologies would be robust and not require substantial fine tuning of kinetic parameters \cite{robustness, alon}. Defining the precise parameter ranges and topological structures which promote stability among these processes may enable one to actually search for these common motifs across a broad range of well-studied metabolic networks. As more data is collected characterizing the rates of reactions and magnitudes of concentrations in metabolism, the possibility of such a search becomes more likely. Searching for dynamical behaviors by varying topologies and parameter ranges is quickly becoming a prototypical problem in systems biology. Work such as \cite{lu, review_inverse} demonstrate that solutions to these problems can have quite fruitful repercussions in our ability to understand naturally occurring biological systems and synthetically design novel genetic and signalling networks.

Finally, the observation that metabolic cycles have robust steady state properties may have ramifications in the study of early metabolism. Several studies have suggested that the early stages of molecular self-organization and self-reproduction may have involved the emergence of self-sustaining proto-metabolic cycles \cite{ganti, morowitz}. Arguments such as those presented in \cite{Orgel}, however, remind us that the earliest biological reactions may have taken place in environments which siphoned off the products of these reactions. In the case of cycles, these dilutions would have potentially precluded a replenishment of substrates in the cycle, and thus prevented a stable steady state. Irrespective of this open question, our work suggests that when autocatalytic cycles did first appear, they would have benefitted from inherent robustness to environmental perturbations. In general, these cycles may be only a single instance of a large array of other metabolic topologies which exhibit equally robust stability properties. Future studies may address whether these stability properties played a role in establishing the core of current cellular metabolism.

\section{Acknowledgements}
The authors would like to thank Erik Sherwood for many thoughtful discussions relating to nonlinear dynamics and Nicola Soranzo for his guidance regarding CRNT. This work was supported by grants from the Office of Science (BER), U.S. Department of Energy (No. DE-FG02-07ER64388 and No. DE-FG02-07ER64483), NASA (NASA Astrobiology Institute, NNA08CN84A), the National Institute of Health (1R01 GM078209) and the National Science Foundation (NSF DMS-0602204 "EMSW21-RTG, BioDynamics at Boston University").

\bibliographystyle{elsarticle-num}
\bibliography{cycle_stability_arxiv}

\appendix
\section{Structural Kinetic Modeling (SKM)}
\label{sec:SKM}
The methods used in this paper are based on the SKM framework \cite{SKM}. SKM is a specific application of generalized modeling \cite{GM} in which normalized parameters replace conventional kinetic parameters (such as $V_{max}$ or $K_{M}$) in the modeling of metabolic networks. The normalized parameters have a direct connection to the original kinetic parameters, but are much easier to work with. As will be shown below, these parameters usually have well-defined and extremely limited ranges (e.g. [0,1]), and sampling them across this range effectively samples all possible values of the original kinetic parametsrs.

The goal of SKM is to capture the local stability properties of a biochemical system. In this sense, it bridges the gap between genome-scale steady state modeling \cite{FBA} and explicit kinetic modeling of a metabolic process. To accomplish this, one usually determines the Jacobian $J$ of the system of interest and evalutes it at its equilibrium. Assuming knowledge of the form of kinetic rate laws for a CRN, it is quite possible to write down the corresponding $J$. However, in many cases $J$ will be unnecessarily complicated and quite difficult to work with. By performing a change of variables, SKM actually simplifies the mathematical form of each entry in $J$. The new entries are in almost all cases easier to work with. Much of this material is taken directly from \cite{SKM} and we refer readers to that reference and its supplementary materials for more information.

If we let \textbf{S} be the \textit{m}-dimensional vector of metabolite concentrations, $\mathbf{N}$ be the $\mathit{m} \times \mathit{r}$ stoichiometric matrix, and $\mathit{v}$ be the \textit{r}-dimensional vector of reaction rates, then we can describe the dynamics of the system with the equation

\begin{equation}
	\frac{d\mathbf{S}}{dt}=\mathbf{N\mathit{v}(S,k)}
\end{equation}

where $v(\mathbf{S,k})$ denotes that the reaction rates are dependent on both metabolite concentrations \textbf{S} and unknown parameters (such as Michaelis-Menten constants) \textbf{k}. If we assume that a non-negative steady state $\textbf{S}^{0}$ exists, then we can redefine our system using the definitions

\begin{equation}	 \mathit{x_{i}}=\frac{\mathit{S_{i}(t)}}{S_{\mathit{i}}^{0}},\mathbf{\Lambda}_{ij}=N_{ij}\frac{\mathit{v_{j}}(\mathbf{S}^{0})}{S^{0}_{i}},\mu_{j}(\mathbf\mathit{x})=\frac{\mathit{v_{j}}(\mathbf{S})}{\mathit{v_{j}}(\mathbf{S}^{0})}
\end{equation}

where $i = 1...m$ and $j = 1...r$. Now, $\mathbf{x}$ is a vector of metabolite concentrations normalized with respect to their steady state concentrations and $\mathbf{\mu}$ represents flux normalized with respect to steady state flux values. The matrix $\mathbf{\Lambda}$ represents the stoichiometric matrix normalized with respect to steady state fluxes and steady state metabolite concentrations. We can now rewrite the system of differential equations as

\begin{equation}
	\frac{d\mathbf{x}}{dt} = \mathbf{\Lambda\mu(x)}
\end{equation}

Now, we can write the Jacobian and evaluate it at $\mathbf{x}^{0} = 1$ (which, because of the way $\mathbf{x}$ is defined, is actually the equilibrium of the system). Calculating the Jacobian, we find that

\begin{equation}
\mathbf{J_{x}} = \mathbf{\Lambda}\frac{\partial{\mathbf{\mu(x)}}}{\delta\mathbf{x}}
\label{eq:Jx}
\end{equation}

If we evaluate (\ref{eq:Jx}) at $\mathbf{x}^{0} = 1$ we find

\begin{equation}
\mathbf{J_{x}} \equiv \mathbf{\Lambda \theta ^{\mu}_{x}}
\end{equation}

Note that the equations above were derived without regard to the actual form of the kinetic equations that determined the ODE system. The matrix $\mathbf{\theta^{\mu}_{x}}$ contains elements which represent the degree of saturation of normalized flux $\mathbf{\mu}_{j}$ with respect to normalized substrate concentration ${\mathbf{x}_{i}}$. In terms of derivatives, each element of $\mathit{\theta}$ represents the degree of change in a flux as a particular metabolite is incrementally increased. This is analogous to the concept of elasticity in metabolic control analysis \cite{MCA}.

What does the $\mathbf{\theta}$ matrix look like? Its columns correspond to each metabolite, and its rows to each flux. A non-zero element $\theta^{i}_{j}$ in the matrix represents the effect a small change in a metabolite $j$ has on flux $i$. In the case of Michaelis-Menten kinetics, this element in the matrix may take values ranging from [0,1]. In the case of standard competitive inhibition (e.g. allosteric inhibition by a product), the element takes values in [-1,0].

To usefully illustrate the meaning of a single $\theta$ element, consider an equation following Michaelis-Menten kinetics shown in (\ref{eq:MMapp}). First, we write $\mu(x)$, which we recall is the flux normalized by the flux at the steady state. Here, $S_0$ is the concentration of the substrate $S$ at steady state. Then, we manipulate the equation by substituting $xS_0$ for $S$, where $x$ is the normalized steady state concentration of substrate $S$. The result is shown in (\ref{eq:MMmu}).

\begin{equation}
V = \frac{k_{2}E_{0}S}{K_{M}+S}
\end{equation}

\begin{equation}
\mu(x) = \frac{\frac{k_{2}E_{0}S}{K_{M}+S}}{\frac{k_{2}E_{0}S_0}{K_{M}+S_0}} = x\frac{K_M + S_0}{K_M + xS_0}
\end{equation}

Finally, we take a derivative with respect to $x$ and evaluate it at $x = 1$ to obtain $\theta$ in (\ref{eq:MMderiv}). Notice that $\theta$ can only take values between 0 and 1 for any positive value of $S_0$.
\begin{equation}
\theta = \frac{1}{1+\frac{S_0}{K_M}}
\end{equation}

It is also possible to approximate the value of $\theta$ for different extremes of saturation. In the regime of extremely low substrate concentrations when the enzyme catalyzing the reaction is not saturated at all, formation of product can be approximated by
\begin{equation}
\frac{dP}{dt} = \frac{k_{2}E_{0}S}{K_{M}+S} \approx \frac{k_{2}E_{0}S}{K_{M}}
\end{equation}

which is approximately linear in $S$. In this case, $\theta^{i}_{j}$ would be approximately 1. This is identically the case for mass-action kinetics, where the rate of product formation is linear with the amount of substrate. In contrast, at extremely high substrate concentrations,

\begin{equation}
\frac{dP}{dt} = \frac{k_{2}E_{0}S}{K_{M}+S} \approx k_{2}E_{0}
\end{equation}

In this case, $S$ has nearly no effect on the rate of product formation, and so $\theta^{i}_{j}$ is approximately 0. For all other substrate concentrations, $\theta$ falls between 0 and 1. Note that the above analysis can be performed for other forms of kinetic equations including mass-action and Hill-type equations. In these cases, the values for $\theta^{i}_{j}$ are similarly constrained. For example, in the case of Hill kinetics, $\theta^{i}_{j}$ falls in the range [0,$n$], where $n$ is the Hill constant. For more information on analyzing other types of kinetics, see the supplementary materials in \cite{SKM}.

The power of SKM comes from the parameterization illustrated above. Each element in $\mathbf{\theta}$ has a precise correspondence to some combination of kinetic parameters in the original model. However, it is far more tractable to study a system using $\theta$ parameters rather than the original kinetic parameters. To see this more clearly, recall that in most cases, biochemical kinetic constants are poorly estimated. In order to build precise ODE models for biochemical systems, it is usually necessary to actually choose values for these constants. While the chosen values may be estimated from experimental measurements, hidden dependencies in the model may actually results in non-obvious correlations between parameters that can strongly affect the output of the model.

On the other hand, if one is only interested in the \textit{stability} of a characterized steady-state, it may not be necessary to actually have knowledge of kinetic parameters. Experimental measurements can provide data on absolute metabolite concentrations and flux values, and the stoichiometry is often known \textit{a priori}. Then, one can parameterize the system as shown above and sample many possible combinations of $\theta$ parameters. For each unique set of $\theta$ parameters, the stability of the Jacobian is determined and recorded. Analysis of the results can lead to several important conclusions such as which $\theta$ parameters have the strongest correlation with stability of the entire metabolic system. Thus, the benefit of employing SKM over other techniques is that it provides the means to analyze and make sense of a large number of possible cases of a metabolic network, rather than a single instance.

Traditionally, SKM has been used as a computational tool to reveal relationships between the stability of a metabolic module and the parameters which govern its kinetics. However, many aspects of the SKM parameterization lend themselves to analytical methods. By using $\theta$ parameters, it is no longer necessary to decide on the kinetic form (e.g. mass action, Michaelis-Menten, Hill) of the chemical reactions taking place. Instead, the problem of studying stability is reduced to finding the signs of the roots of a polynomial equation (the characterstic equation of the Jacobian $J$), whose coefficients are determined by the fluxes, absolute metabolite concentrations, and $\theta$'s of the system at steady state. Moreover, it is in many cases possible to limit the values of $\theta$ (e.g.- if simple Michaelis-Menten kinetics are assumed without inhibition, $\theta$ falls between 0 and 1). If the signs of the roots are all negative, then the eigenvalues of the Jacobian are negative and the current steady state is stable.

At first glance, it is unclear whether this will make things easier or harder; values for fluxes and concentrations are almost never known. However, our work shows that the absence of this knowledge does not make the problem of stability intractable. Instead, we operate with the following idea: we assume that the steady state flux vector $v$ and the metabolite concentration vector $X$ are variables, and we do not specify them. Instead, we simply determine the stability of the system in terms of these variables. Then, we can find trends in stability as the flux and metabolite concentrations change: it may be that as a concentration goes up, the stability of the system tends to fall. Isolating these trends is the motivation of our analysis.

It is worthwhile to clarify what it means to change a steady-state parameter like a flux or a $\theta$. At first glance, it may not be obvious that it is even possible to alter such a parameter. In other words, suppose that at steady state, the concentration of metabolite $A$ in some network is 1. What does it mean to increase the concentration of $A$? If we do so without altering any other parameters, the metabolic network will in all likelihood not be in steady state any longer. However, there may exist (and in fact for the systems described in this paper there does exist) another set of concentrations, fluxes, and $\theta$'s which does lead to a steady state at the increased concentration of $A$. The problem, again, is that we do not know their actual values. However, we may be able to infer, based on changes in the coefficients of the characteristic equation, that if we were to sample a large number of unique instances of the Jacobian for both concentrations of $A$, the higher concentration may actually be \textit{more likely} to be stable. This is in fact the method we will use to study autocatalytic cycles later in the paper. However, in the next two sections, we will demonstrate that for certain types of cycles, it is not necessary to go to such great lengths in order to characterize stability; it will come much more naturally.

\section{Linear Chains}
\label{sec:LinearChain}
Up to this point, linear metabolic networks have been ignored. The reason for this is that, using the same analysis as above, it is trivial to show such a network is stable. In this appendix, we provide a short example illustrating this point. Consider the metabolic network shown in Figure 3 with flux $F$ flowing through it and observed steady states $(A^0 B^0)$.

\begin{figure}[h]
	\begin{center}
		\includegraphics{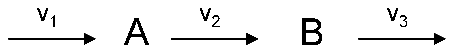}
	\end{center}
	\caption{A linear chain}
	\label{fig:linear}
\end{figure}

If we follow the same analysis as before (writing $\mathbf{\Lambda}$ and $\mathbf{\theta}$ and calculating the Jacobian $J$), we will find

\begin{equation}
J
=\left[
\begin{array}{cc}
\frac{-F}{A^0} & 0\\
\vspace{1pt} \\
\frac{F}{B^0} & \frac{-F}{B^0}
\end{array}
\right]
\end{equation}

The eigenvalues of the Jacobian are then $\frac{-F}{A^0}$ and $\frac{-F}{B^0}$, both of which are negative. Therefore, the system is
stable.

\section{Cofactors}
\label{sec:Cofactors}
The reaction system shown below is the complete, more general version of the simple system analyzed above.

Now, we can write a system very similar to the one in Section 2. Consider the example shown in the figure below.
\begin{figure}[h]
	\begin{center}
		\includegraphics{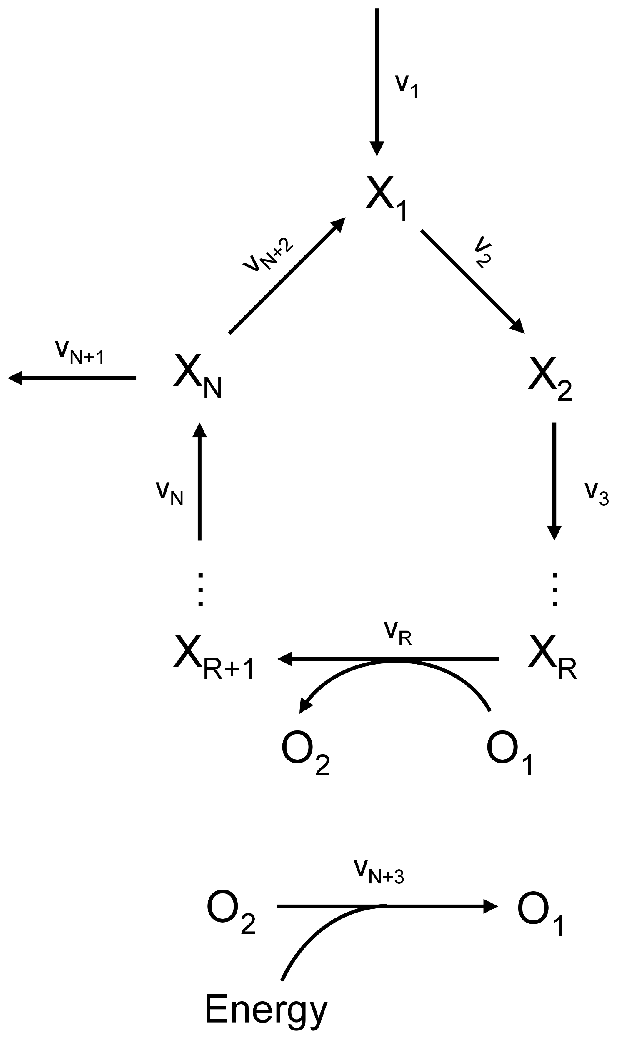}
	\end{center}
	\caption{A complete metabolic cycle with cofactors}
\end{figure}

The generalized form of $\mathbf{\Lambda}$ and $\mathbf{\theta}$ is shown below. Note that we make no further assumptions about the steady states of the system: $X^0 = (X_1^0, X_2^0,\dots,X_N^0)$ and the flux through the system remains equal to $F$. With $N$ metabolites and $N+3$ reactions, $\Lambda$ is $N \times (N+3)$ and $\theta$ is $(N+3) \times N$.

Note that because the cofactors come in a conserved pair, one of them can be removed from the system. This modifies the entry in the bottom right element of the Jacobian.

\begin{landscape}
\begin{center}
$\mathbf{\Lambda} = \left[
\begin{array}{cccccccccccccc}
\frac{\alpha F}{X_1^0} & \frac{-F}{X_1^0} & 0 & \cdots & 0 & 0 & 0 & 0 & \cdots & 0 & 0 & \frac{(1-\alpha)}{X_1^0} & 0\\
\vspace{1pt} \\
0 & \frac{F}{X_2^0} & \frac{-F}{X_2^0} & \cdots & 0 & 0 & 0 & 0 & \cdots & 0 & 0 & 0 & 0\\
\vspace{1pt} \\
0 & 0 & \frac{F}{X_3^0} & \cdots & 0 & 0 & 0 & 0 &\cdots & 0 & 0 & 0 & 0\\
\vspace{1pt} \\
\vdots & \vdots & \vdots & & \vdots & \vdots & \vdots & \vdots & & \vdots & \vdots & \vdots \\
\vspace{1pt}\\
0 & 0 & 0 & \cdots & \frac{F}{X_{R-1}^0} & \frac{-F}{X_{R-1}^0} & 0 & 0 & \cdots & 0 & 0 & 0 & 0 \\
\vspace{1pt}\\
0 & 0 & 0 & \cdots & 0 & \frac{F}{X_{R}^0} & \frac{-F}{X_{R}^0} & 0 & \cdots & 0 & 0 & 0 & 0\\
\vspace{1pt}
0 & 0 & 0 & \cdots & 0 & 0 & \frac{F}{X_{R+1}^0} & \frac{-F}{X_{R+1}^0} & \cdots & 0 & 0 & 0 & 0\\
\vspace{1pt}
0 & 0 & 0 & \cdots & 0 & 0 & 0 & \frac{F}{X_{R+2}^0} & \cdots & 0 & 0 & 0 & 0 \\
\vspace{1pt}
\vdots & \vdots & \vdots & & \vdots & \vdots & \vdots & & \vdots & \vdots & \vdots \\
\vspace{1pt}
0 & 0 & 0 & \cdots &  0 & 0 & 0 & 0 & \cdots & \frac{-F}{X_{N-1}^0} & 0 & 0 & 0 \\
\vspace{1pt}
0 & 0 & 0 & \cdots & 0 & 0 & 0 & 0 & \cdots & \frac{F}{X_{N}^0} & \frac{-\alpha F}{X_N^0} & \frac{(1-\alpha)F}{X_N^0} & 0 \\
\vspace{1pt}
0 & 0 & 0 & \cdots & 0 & 0 & \frac{-F}{O_1^0} & 0 & \cdots & 0 & 0 & 0 & \frac{F}{O_1^0}
\end{array}
\right]$
\end{center}

and
\begin{center}
$\mathbf{\theta} = \left[
\begin{array}{cccccccc}
0 & 0 & 0 & \cdots & 0 & \cdots & 0 & 0 \\
\theta_1 & 0 & 0 & \cdots & 0 & \cdots & 0 & 0 \\
0 & \theta_2 & 0 & \cdots & 0 & \cdots & 0 & 0 \\
0 & 0 & \theta_3 & \cdots & 0 & \cdots & 0 & 0 \\
\vdots & \vdots & \vdots & \ddots & \vdots && \vdots & \vdots \\
0 & 0 & 0 & \cdots & \theta_R & \cdots & 0 & \theta_{N+2} \\
\vdots & \vdots & \vdots & & \vdots & \ddots & \vdots & \vdots \\
0 & 0 & 0 & \cdots & 0 & \cdots & \theta_N & 0\\
0 & 0 & 0 & \cdots & 0 & \theta_{N+1} & 0\\
0 & 0 & 0 & \cdots & 0 & \cdots & 0 & -\frac{\theta_{N+3}O_1^0}{O_2^0}
\end{array}
\right]$
\end{center}
Note that the last row of $\Lambda$ is omitted because the cofactors come as a conserved pair. Furthermore, note that the bottom right element of $\theta$ is replaced with a negative element in order to account for this conservation. Aftering factor out $F$, the Jacobian becomes

$\left[
\begin{array}{cccccccccccccc}
\frac{-\theta_1}{X_1^0} & 0 & 0 & \cdots & 0 & 0 & 0 & 0 & \cdots & 0 & \frac{(1-\alpha)\theta_{N+1}}{X_1^0} & 0\\
\vspace{1pt} \\
\frac{\theta_1}{X_2^0} & \frac{-\theta_2}{X_2^0} & 0 & \cdots & 0 & 0 & 0 & 0 & \cdots & 0 & 0 & 0\\
\vspace{1pt} \\
0 & \frac{\theta_2}{X_3^0} & \frac{-\theta_3}{X_3^0} & \cdots & 0 & 0 & 0 & 0 &\cdots & 0 & 0 & 0\\
\vspace{1pt} \\
\vdots & \vdots & \vdots & & \vdots & \vdots & \vdots & \vdots & & \vdots & \vdots & \vdots \\
\vspace{1pt}\\
0 & 0 & 0 & \cdots & \frac{-\theta_{R-1}}{X_{R-1}^0} & 0 & 0 & 0 & \cdots & 0 & 0 & 0 \\
\vspace{1pt}\\
0 & 0 & 0 & \cdots & \frac{\theta_{R-1}}{X_{R}^0} & \frac{-\theta_{R}}{X_{R}^0} & 0 & 0 & \cdots & 0 & 0 & \frac{-\theta_{N+2}}{X_{R}^0}\\
\vspace{1pt}
0 & 0 & 0 & \cdots & 0 & \frac{\theta_R}{X_{R+1}^0} & \frac{-\theta_{R+1}}{X_{R+1}^0} & 0 & \cdots & 0 & 0 & \frac{\theta_{N+2}}{X_{R+1}^0}\\
\vspace{1pt}
0 & 0 & 0 & \cdots & 0 & 0 & \frac{\theta_{R+1}}{X_{R+2}^0} & \frac{-\theta_{R+2}}{X_{R+2}^0} & \cdots & 0 & 0 & 0 \\
\vspace{1pt}
\vdots & \vdots & \vdots & & \vdots & \vdots & \vdots & & \vdots & \vdots & \vdots \\
\vspace{1pt}
0 & 0 & 0 & \cdots &  0 & 0 & 0 & 0 & \cdots & \frac{-\theta_{N-1}}{X_{N-1}^0} & 0 & 0 \\
\vspace{1pt}
0 & 0 & 0 & \cdots & 0 & 0 & 0 & 0 & \cdots & \frac{\theta_{N-1}}{X_N^0} & \frac{-\alpha \theta_N - (1-\alpha )\theta_{N+1}}{X_N^0} & 0 \\
\vspace{1pt}
0 & 0 & 0 & \cdots & 0 & \frac{-\theta_R}{O_1^0} & 0 & 0 & \cdots & 0 & 0 & \frac{-\theta_{N+2}}{O_1^0} - \frac{-\theta_{N+3}}{O_2^0}
\end{array}
\right]$

Next, we evaluate take the determinant $|J - \lambda I|$ and take the Laplace expansion over the last column. This yields three smaller determinants. The first is $\frac{-\theta_{N+2}}{O_1^0} - \frac{\theta_{N+3}}{O_2^0}$ multiplied by the determinant of:

$\left[
\begin{array}{ccccccccccccc}
\frac{-\theta_1}{X_1^0}-\lambda & 0 & 0 & \cdots & 0 & 0 & 0 & 0 & \cdots & 0 & \frac{(1-\alpha)\theta_{N+1}}{X_1^0}\\
\vspace{1pt} \\
\frac{\theta_1}{X_2^0} & \frac{-\theta_2}{X_2^0}-\lambda & 0 & \cdots & 0 & 0 & 0 & 0 & \cdots & 0 & 0\\
\vspace{1pt} \\
0 & \frac{\theta_2}{X_3^0} & \frac{-\theta_3}{X_3^0}-\lambda & \cdots & 0 & 0 & 0 & 0 &\cdots & 0 & 0\\
\vspace{1pt} \\
\vdots & \vdots & \vdots & & \vdots & \vdots & \vdots & \vdots & & \vdots & \vdots\\
\vspace{1pt}\\
0 & 0 & 0 & \cdots & \frac{-\theta_{R-1}}{X_{R-1}^0}-\lambda & 0 & 0 & 0 & \cdots & 0 & 0\\
\vspace{1pt}\\
0 & 0 & 0 & \cdots & \frac{\theta_{R-1}}{X_{R}^0} & \frac{-\theta_{R}}{X_{R}^0}-\lambda & 0 & 0 & \cdots & 0 & 0\\
\vspace{1pt}
0 & 0 & 0 & \cdots & 0 & \frac{\theta_R}{X_{R+1}^0} & \frac{-\theta_{R+1}}{X_{R+1}^0}-\lambda & 0 & \cdots & 0 & 0\\
\vspace{1pt}
0 & 0 & 0 & \cdots & 0 & 0 & \frac{\theta_{R+1}}{X_{R+2}^0} & \frac{-\theta_{R+2}}{X_{R+2}^0}-\lambda & \cdots & 0 & 0 \\
\vspace{1pt}
\vdots & \vdots & \vdots & & \vdots & \vdots & \vdots & \vdots & & \vdots & \vdots\\
\vspace{1pt}
0 & 0 & 0 & \cdots &  0 & 0 & 0 & 0 & \cdots & \frac{-\theta_{N-1}}{X_{N+1}^0}-\lambda & 0 \\
\vspace{1pt}
0 & 0 & 0 & \cdots & 0 & 0 & 0 & 0 & \cdots & \frac{\theta_{N-1}}{X_N^0} & \frac{-\alpha \theta_N - (1-\alpha )\theta_{N+1}}{X_N^0}-\lambda
\end{array}
\right]$

This determinant is easily evaluated and results in
\begin{equation}
\left(\prod_{i=1}^{N-1}{\left(-\lambda - \frac{F\theta_i}{X_i^0}\right)}\right)\left(\frac{-\lambda - \alpha F\theta_N - (1- \alpha)F\theta_{N+1}}{X_N^0}\right)+ (-1)^{N-1}\frac{(1-\alpha)\theta_{N+1}}{X_1^0}\prod_{i=1}^{N-1}\frac{\theta_i}{X_{i+1}^0}
\end{equation}

The second matrix resulting from the Laplace expansion of the original Jacobian is $(-1)^{N+R+1}\frac{-\theta_{N+2}}{X_R^0}$ multiplied by the determinant of

$\left[
\begin{array}{ccccccccccccc}
\frac{-\theta_1}{X_1^0}-\lambda & 0 & 0 & \cdots & 0 & 0 & 0 & 0 & \cdots & 0 & \frac{(1-\alpha)\theta_{N+1}}{X_1^0}\\
\vspace{1pt} \\
\frac{\theta_1}{X_2^0} & \frac{-\theta_2}{X_2^0}-\lambda & 0 & \cdots & 0 & 0 & 0 & 0 & \cdots & 0 & 0\\
\vspace{1pt} \\
0 & \frac{\theta_2}{X_3^0} & \frac{-\theta_3}{X_3^0}-\lambda & \cdots & 0 & 0 & 0 & 0 &\cdots & 0 & 0\\
\vspace{1pt} \\
\vdots & \vdots & \vdots & & \vdots & \vdots & \vdots & \vdots & & \vdots & \vdots \\
\vspace{1pt}\\
0 & 0 & 0 & \cdots & \frac{-\theta_{R-1}}{X_{R-1}^0}-\lambda & 0 & 0 & 0 & \cdots & 0 & 0\\
\vspace{1pt}
0 & 0 & 0 & \cdots & 0 & \frac{\theta_R}{X_{R+1}^0} & \frac{-\theta_{R+1}}{X_{R+1}^0}-\lambda & 0 & \cdots & 0 & 0\\
\vspace{1pt}
0 & 0 & 0 & \cdots & 0 & 0 & \frac{\theta_{R+1}}{X_{R+2}^0} & \frac{-\theta_{R+2}}{X_{R+2}^0}-\lambda & \cdots & 0 & 0\\
\vspace{1pt}
\vdots & \vdots & \vdots & & \vdots & \vdots & \vdots & \vdots & & \vdots & \vdots \\
\vspace{1pt}
0 & 0 & 0 & \cdots &  0 & 0 & 0 & 0 & \cdots & \frac{-\theta_{N-1}}{X_{N+1}^0}-\lambda & 0 \\
\vspace{1pt}
0 & 0 & 0 & \cdots & 0 & 0 & 0 & 0 & \cdots & \frac{\theta_{N-1}}{X_N^0} & \frac{-\alpha \theta_N - (1-\alpha )\theta_{N+1}}{X_N^0}-\lambda\\
\vspace{1pt}
0 & 0 & 0 & \cdots & 0 & \frac{-\theta_R}{O_1^0} & 0 & 0 & \cdots & 0 & 0
\end{array}
\right]$

If another Laplace expansion is now taken along the $N^{th}$ row, then the resulting matrix is

$\left[
\begin{array}{cccccccccccc}
\frac{-\theta_1}{X_1^0}-\lambda & 0 & 0 & \cdots & 0 & 0 & 0 & \cdots & 0 & \frac{(1-\alpha)\theta_{N+1}}{X_1^0}\\
\vspace{1pt} \\
\frac{\theta_1}{X_2^0} & \frac{-\theta_2}{X_2^0}-\lambda & 0 & \cdots & 0 & 0 & 0 & \cdots & 0 & 0\\
\vspace{1pt} \\
0 & \frac{\theta_2}{X_3^0} & \frac{-\theta_3}{X_3^0}-\lambda & \cdots & 0 & 0 & 0 &\cdots & 0 & 0\\
\vspace{1pt} \\
\vdots & \vdots & \vdots & & \vdots & \vdots & \vdots & & \vdots & \vdots \\
\vspace{1pt}\\
0 & 0 & 0 & \cdots & \frac{-\theta_{R-1}}{X_{R-1}^0}-\lambda & 0 & 0 & \cdots & 0 & 0\\
\vspace{1pt}
0 & 0 & 0 & \cdots & 0 & \frac{-\theta_{R+1}}{X_{R+1}^0}-\lambda & 0 & \cdots & 0 & 0\\
\vspace{1pt}
0 & 0 & 0 & \cdots & 0 & \frac{\theta_{R+1}}{X_{R+2}^0} & \frac{-\theta_{R+2}}{X_{R+2}^0}-\lambda & \cdots & 0 & 0\\
\vspace{1pt}
\vdots & \vdots & \vdots & & \vdots & \vdots & \vdots & & \vdots & \vdots \\
\vspace{1pt}
0 & 0 & 0 & \cdots &  0 & 0 & 0 & \cdots & \frac{-\theta_{N-1}}{X_{N+1}^0}-\lambda & 0 \\
\vspace{1pt}
0 & 0 & 0 & \cdots & 0 & 0 & 0 & \cdots & \frac{\theta_{N-1}}{X_N^0} & \frac{-\alpha \theta_N - (1-\alpha )\theta_{N+1}}{X_N^0}-\lambda\\
\end{array}
\right]$

The resulting term for this determinant is

\begin{equation}
\left( \frac{-\alpha \theta _N - (1-\alpha )\theta_{N+1}}{X_N^0} - \lambda \right)\prod^{N-1}_{i=1,i\neq R} {\left(\frac{-\theta_i}{X_{i}^0}-\lambda \right)}
\end{equation}

The third matrix resulting from the Laplace expansion is $(-1)^{N+R+2} \frac{\theta_{N+2}}{X_{R+1}^0}$ multiplied by the determinant of

$\left[
\begin{array}{ccccccccccccc}
\frac{-\theta_1}{X_1^0}-\lambda & 0 & 0 & \cdots & 0 & 0 & 0 & 0 & \cdots & 0 & \frac{(1-\alpha)\theta_{N+1}}{X_1^0}\\
\vspace{1pt} \\
\frac{\theta_1}{X_2^0} & \frac{-\theta_2}{X_2^0}-\lambda & 0 & \cdots & 0 & 0 & 0 & 0 & \cdots & 0 & 0\\
\vspace{1pt} \\
0 & \frac{\theta_2}{X_3^0} & \frac{-\theta_3}{X_3^0}-\lambda & \cdots & 0 & 0 & 0 & 0 &\cdots & 0 & 0\\
\vspace{1pt} \\
\vdots & \vdots & \vdots & & \vdots & \vdots & \vdots & \vdots & & \vdots & \vdots \\
\vspace{1pt}\\
0 & 0 & 0 & \cdots & \frac{-\theta_{R-1}}{X_{R-1}^0}-\lambda & 0 & 0 & 0 & \cdots & 0 & 0 \\
\vspace{1pt}\\
0 & 0 & 0 & \cdots & \frac{\theta_{R-1}}{X_{R}^0} & \frac{-\theta_{R}}{X_{R}^0}-\lambda & 0 & 0 & \cdots & 0 & 0\\
\vspace{1pt}
0 & 0 & 0 & \cdots & 0 & 0 & \frac{\theta_{R+1}}{X_{R+2}^0} & \frac{-\theta_{R+2}}{X_{R+2}^0}-\lambda & \cdots & 0 & 0\\
\vspace{1pt}
\vdots & \vdots & \vdots & & \vdots & \vdots & \vdots & \vdots & & \vdots & \vdots \\
\vspace{1pt}
0 & 0 & 0 & \cdots &  0 & 0 & 0 & 0 & \cdots & \frac{-\theta_{N-1}}{X_{N+1}^0}-\lambda & 0 \\
\vspace{1pt}
0 & 0 & 0 & \cdots & 0 & 0 & 0 & 0 & \cdots & \frac{\theta_{N-1}}{X_N^0} & \frac{-\alpha \theta_N - (1-\alpha )\theta_{N+1}}{X_N^0}-\lambda\\
\vspace{1pt}
0 & 0 & 0 & \cdots & 0 & \frac{-\theta_R}{O_1^0} & 0 & 0 & \cdots & 0 & 0
\end{array}
\right]$

Taking another Laplace expansion over the $N^{th}$ column, then the resulting matrix is

$\left[
\begin{array}{ccccccccccccc}
\frac{-\theta_1}{X_1^0}-\lambda & 0 & 0 & \cdots & 0 & 0 & 0 & \cdots & 0 & \frac{(1-\alpha)\theta_{N+1}}{X_1^0}\\
\vspace{1pt} \\
\frac{\theta_1}{X_2^0} & \frac{-\theta_2}{X_2^0}-\lambda & 0 & \cdots & 0 & 0 & 0 & \cdots & 0 & 0\\
\vspace{1pt} \\
0 & \frac{\theta_2}{X_3^0} & \frac{-\theta_3}{X_3^0}-\lambda & \cdots & 0 & 0 & 0 &\cdots & 0 & 0\\
\vspace{1pt} \\
\vdots & \vdots & \vdots & & \vdots & \vdots & \vdots & & \vdots & \vdots \\
\vspace{1pt}\\
0 & 0 & 0 & \cdots & \frac{-\theta_{R-1}}{X_{R-1}^0}-\lambda & 0 & 0 & \cdots & 0 & 0 \\
\vspace{1pt}\\
0 & 0 & 0 & \cdots & \frac{\theta_{R-1}}{X_{R}^0} &  0 & 0 & \cdots & 0 & 0\\
\vspace{1pt}
0 & 0 & 0 & \cdots & 0 & \frac{\theta_{R+1}}{X_{R+2}^0} & \frac{-\theta_{R+2}}{X_{R+2}^0}-\lambda & \cdots & 0 & 0\\
\vspace{1pt}
\vdots & \vdots & \vdots & & \vdots & \vdots & \vdots & & \vdots & \vdots \\
\vspace{1pt}
0 & 0 & 0 & \cdots &  0 & 0 & 0 & \cdots & \frac{-\theta_{N-1}}{X_{N+1}^0}-\lambda & 0 \\
\vspace{1pt}
0 & 0 & 0 & \cdots & 0 & 0 & 0 & \cdots & \frac{\theta_{N-1}}{X_N^0} & \frac{-\alpha \theta_N - (1-\alpha )\theta_{N+1}}{X_N^0}-\lambda\\
\end{array}
\right]$

The resulting term for this determinant is

\begin{equation}
(-1)^{N}\frac{(1-\alpha)\theta_{N+1}}{X_1^0}\prod^{N-1}_{i=1,i\neq R}{\frac{\theta_i}{X_{i+1}^0}}
\end{equation}

Now we can sum all terms to get

\begin{eqnarray}
\left( \frac{-\theta_{N+2}}{O_1^0} - \frac{\theta_{N+3}}{O_2^0} - \lambda \right) \left( \prod_{i=1}^{N-1} { \left(-\lambda - \frac{F\theta_i}{X_i^0}\right)} \left(-\lambda - \frac{ \alpha \theta_N + (1- \alpha) \theta_{N+1}}{X_N^0} \right)+ (-1)^{N-1}\frac{(1-\alpha) \theta_{N+1}}{X_1^0}\prod_{i=1}^{N-1}\frac{\theta_i}{X_{i+1}^0}\right) + \nonumber \\
+ (-1)^{2N+2R+1}\frac{\theta_R \theta_{N+2}}{O_1^0X_R^0}\left( \frac{-\alpha \theta _N - (1-\alpha )\theta_{N+1}}{X_N^0} - \lambda \right)\prod^{N-1}_{i=1,i\neq R} {\left(\frac{-\theta_i}{X_{i}^0}-\lambda \right)} + \nonumber \\
(-1)^{2N+2R+3} \frac{\theta_{R} \theta_{N+2}}{O_1^0X_{R+1}^0}(-1)^{N}\frac{(1-\alpha)\theta_{N+1}}{X_1^0}\prod^{N-1}_{i=1,i\neq R}{\frac{\theta_i}{X_{i+1}^0}}
\end{eqnarray}

The terms on the bottom two lines will cancel with terms from the top two lines, resulting in a polynomial equation of order $N+1$ containing all nonzero coefficients with exactly the same sign. Furthermore, a constant $\lambda^0$ term will remain, indicating that there is no zero eigenvalue. 
\end{landscape}

\end{document}